# Efficient Separation of the Isomeric State $^{26\text{m}}\text{Al}$ from the Intense Ground State Background via Sequential Optical Pumping in Collinear Laser Spectroscopy

**Hoon Yu, Jung Bog Kim, Cheolmin Ham, and Sung Jong Park**

**Abstract**

We propose a novel, highly efficient method for isolating the isomeric state $^{26\text{m}}\text{Al}$ from an overwhelming ground-state $^{26\text{g}}\text{Al}$ background (isomeric ratio $\sim 20:1$) using optical pumping through a 2.0 m flight zone in collinear laser spectroscopy (CLS). To investigate the underlying optical pumping (OP) dynamics, we developed a comprehensive rate equation framework. While the transition pathways can be intuitively conceptualized via a primary 7-manifold scheme, our numerical simulation solves the full 47-level rate equations by explicitly accounting for all degenerate Zeeman sublevels ($m_F$) to rigorously incorporate polarization selection rules and Clebsch-Gordan coefficients. When the continuous acceleration voltage matches the resonance conditions of the $^{26\text{g}}\text{Al}$ hyperfine transitions, the ground-state atoms undergo a 100% efficient transition into uncoupled dark states within the 2.0 m flight zone. Consequently, background fluorescence from the ground state is completely suppressed in the detection chamber, whereas $^{26\text{m}}\text{Al}$ atoms utilize a closed cycling structure to survive the flight zone, yielding a high-intensity, background-free resonance peak.

## 1. Introduction

The study of the exotic odd-odd nucleus Aluminum-26 ($^{26}\text{Al}$) provides crucial insights into stellar nucleosynthesis and galactic $\gamma$-ray astronomy [1]. In particular, the co-existence of the ground state ($^{26\text{g}}\text{Al}$, $I^\pi=5^+$, $T_{1/2}=7.17 \times 10^5\text{ yr}$) and its low-lying isomeric state ($^{26\text{m}}\text{Al}$, $I^\pi=0^+$, $T_{1/2}=6.35\text{ s}$) serves as a key diagnostic tool for understanding thermalization processes in massive stars [1, 2]. Beyond its astrophysical significance, the superallowed $0^+

\rightarrow 0^+$ $\beta^+$ decay of the short-lived metastable state $^{26\text{m}}\text{Al}$ has drawn significant attention from nuclear physicists [3], providing a critical and stringent experimental test for both the conserved-vector-current (CVC) hypothesis and the unitarity of the Cabibbo-Kobayashi-Maskawa (CKM) quark-mixing matrix [3].

To refine the statistical frameworks governing the $V_{ud}$ matrix element, conventional collinear laser spectroscopy (CLS) has long been deployed as a cornerstone technique, as demonstrated in recent high-precision determinations of nuclear charge radii and electromagnetic moments across the nuclear chart [1, 4, 5]. However, while advanced radioactive ion beam facilities globally utilize these optical methods to probe such exotic structures [5, 6], achieving purely background-free measurements of the isomeric state remains a formidable challenge due to the overwhelming presence of the dominant ground-state contaminants.

These high-precision studies underscore the growing necessity for high-purity, background-free isomeric beams. However, extracting pure $^{26m}Al$ optical signals and resolving its transitions using conventional collinear laser spectroscopy (CLS) remains a formidable challenge. In typical isotope separator online (ISOL) facilities, the production mechanism heavily favors the ground state, yielding an overwhelming background of $^{26g}Al$. This typically results in an adverse isomeric ratio ($^{26g}Al$ to $^{26m}Al$) that often exceeds 20:1, as benchmarked by dedicated isomeric beam development studies [7]. Advanced resonance ionization spectroscopy (RIS) setups have demonstrated impressive capabilities in element- and state-selective laser ionization. Nonetheless, achieving complete, background-free suppression within the optical manipulation zone of CLS remains a critical bottleneck. Recently, Plattner et al. successfully overcame the challenge of complete spectral overlap by introducing a groundbreaking approach that leverages their differing half-lives to extract the radius of $^{26m}Al$ [3]. Yet, a general method for real-time optical purification during flight is still highly demanded.

To resolve this limitation, this work proposes a highly efficient state-purification framework to separate $^{26m}Al$ from the intense $^{26g}Al$ background utilizing in-flight optical pumping (OP) within a 2.0 m interaction length. By exploiting the distinct hyperfine structures and transition pathways—benchmarked against high-resolution ultraviolet laser spectroscopy of aluminum atoms—we theoretically demonstrate that the ground-state population can be completely driven into "dark" states with near 100% efficiency. This leaves the isomeric state intact for high-contrast fluorescence detection. The proposed architecture also simplifies the experimental layout by utilizing identical circular polarization ($\sigma\pm$) across both the preparation and detection regions, offering a robust and practical pathway for next-generation ISOL facilities such as RAON.

## 2. Physical Mechanism and Unified Architecture

The isolation mechanism relies on the structural differences in the hyperfine structures (HFS) between the two states. The ground state $^{26\text{g}}\text{Al}$ ($I = 5$) exhibits substantial hyperfine splitting in both the lower $3s^23p~^2\text{P}_{1/2, 3/2}$ and upper $3s^24s~^2\text{S}_{1/2}$ manifolds. Conversely, the isomer $^{26\text{m}}\text{Al}$ ($I = 0$) exhibits no hyperfine splitting due to its zero nuclear spin, resulting in a single, un-split resonance line located within the HFS valleys of the ground state. In the absence of an external magnetic field, each hyperfine level $F$ (or fine structure level $J$) remains exactly $(2F + 1)$-fold (or $(2J + 1)$-fold) degenerate with respect to its magnetic sublevels $m_F$ (or $m_J$).

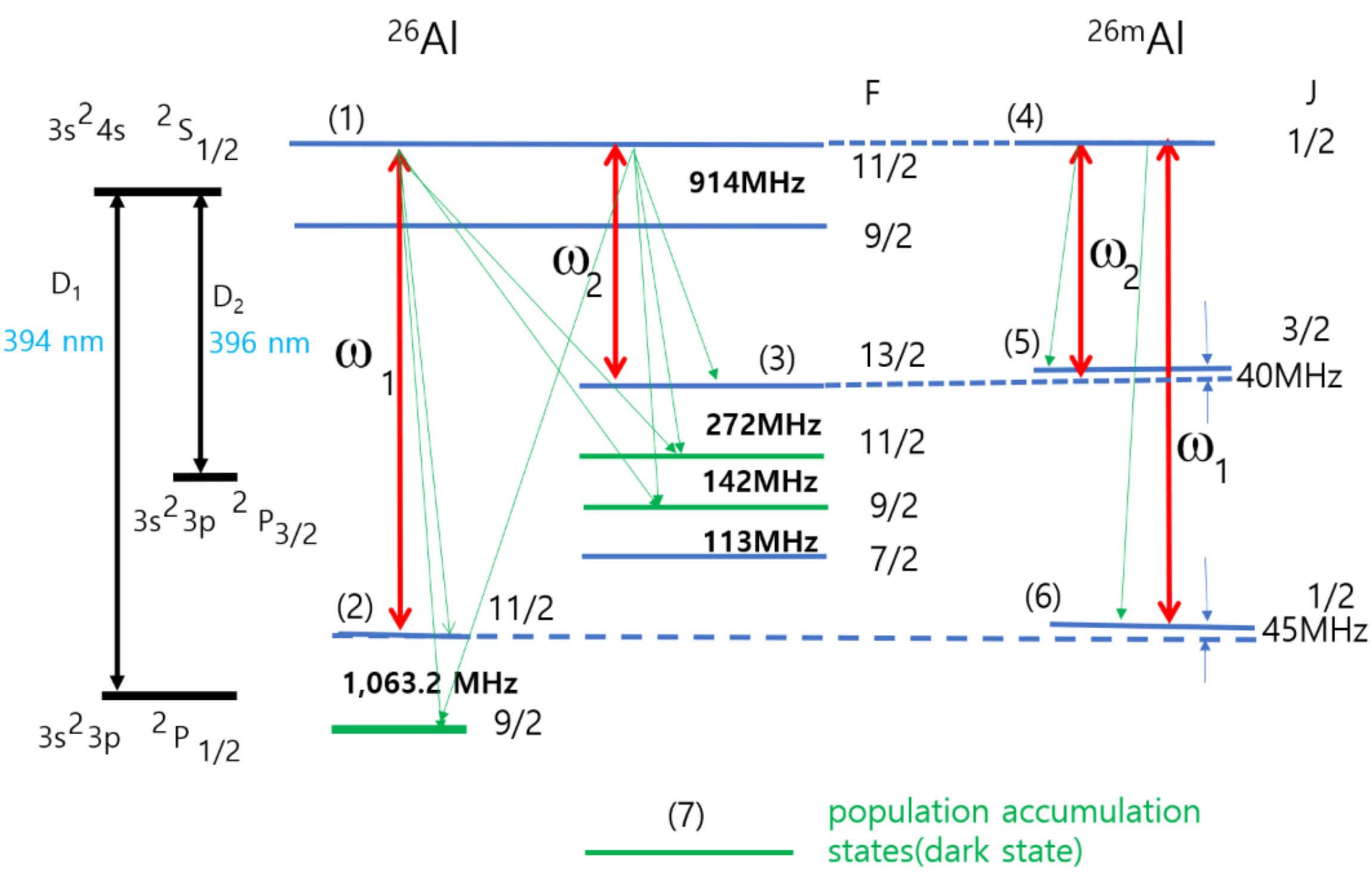

***Fig. 1.*** *Unified energy level diagram and schematic of the 47-level optical pumping and detection channels for $^{26}\text{Al}\,(I = 5)$ and $^{26\text{m}}\text{Al}\,(I = 0)$. The numerical labels (1) through (7) denote specific fine and hyperfine energy levels. Levels (2) and (3) are the driven ground-state hyperfine components simultaneously pumped to the excited state (1) by two distinct laser fields ($\omega_1$ and $\omega_2$), which are systematically depleted into the unified dark state (7) via spontaneous emission. Labels (4), (5), and (6) trace the*

*closed cycling transitions of the isomeric state under the same dual-laser configuration, preserving its population within a 2.0 m flight zone for CLS detection*.

As illustrated in Fig. 1, the total atomic manifold is rigorously categorized into 7 discrete energy levels to elucidate the underlying state dynamics. Labels (2) and (3) represent the specific ground-state hyperfine-splitting levels ($F = 11/2$ and $F = 13/2$, respectively) of the ground state ($^{26\text{g}}\text{Al}$, $I = 5$), which are actively and simultaneously driven by the dual-laser configuration ($\omega_1$ and $\omega_2$ with $\sigma^+$ polarization) up to the excited hyperfine level (1) ($F' = 11/2$ of the $2\text{S}_{1/2}$ manifold).

From this excited state (1), a major fraction of the population spontaneously decays via prominent branching channels exclusively into the lower-lying green-indicated hyperfine states. In contrast, the un-driven $F = 7/2$ level remains completely unaffected by both the laser fields and the spontaneous decay from level (1) due to quantum selection rules, maintaining its initial population throughout the process. Crucially, all these extra states outside the driven transitions—including the green population accumulation states and the isolated $F = 7/2$ level—are completely off-resonance and uncoupled from the two applied laser frequencies ($\omega_1, \omega_2$). To optimize computational efficiency and maintain physical clarity in our 47-level rate equation simulation, these non-interacting states are collectively grouped and modeled as a single, unified surplus manifold represented by Level (7). Under continuous laser driving and $\sigma^+$ selection rules, the population in the active ground states (2) and (3) is systematically swept and irreversibly accumulated into the dark sublevels of Level (7), ensuring the complete in-flight purification of the atomic beam within the 2.0 m interaction length.

In contrast, Labels (4), (5), and (6) describe the un-split electronic levels of the isomeric state ($^{26\text{m}}\text{Al}$, $I = 0$). Since the isomer possesses a zero nuclear spin, these levels are free from ground-state-like hyperfine branching losses and retain their intrinsic $(2J + 1)$-fold magnetic degeneracy. Under the identical $\sigma^+$ dual-laser configuration, the isomer forms a closed cycling transition pathway among these sublevels—specifically driving transitions between levels (6) to (4) via $\omega_1$ and levels (5) to (4) via $\omega_2$. Because there are no alternative decay channels available outside this driven manifold, the isomer preserves its total population through population redistribution among its degenerate magnetic sublevels. This enables the isomer to generate a high-contrast resonance fluorescence signal for collinear laser spectroscopy (CLS) without falling into any uncoupled dark states

## 3. Theoretical Model and 47-Level Rate Equations

To quantitatively evaluate the isolation efficiency based on the architecture described in Section 2, we construct a comprehensive time-dependent rate equation matrix. While the physical mechanism can be intuitively understood via the simplified 7-level scheme (Fig. 1), a rigorous quantitative prediction requires solving the complete magnetic manifolds that directly participate in the dynamics. To implement this, our numerical simulation explicitly incorporates all relevant degenerate Zeeman sublevels ($m_F$ or $m_J$) of the $3s^23p~^2\text{P}_{1/2, 3/2}$ and $3s^24s~^2\text{S}_{1/2}$ states, optimizing the matrix dimension without losing physical rigor.

For the ground state ($^{26\text{g}}\text{Al}$), the system accounts for exactly 39 independent channels: 12 magnetic sublevels for the excited hyperfine level (1) ($F' = 11/2$), 12 sublevels for the driven ground hyperfine level (2) ($F = 11/2$), and 14 sublevels for the driven ground hyperfine level (3) ($F = 13/2$). To track the optical pumping loss, all other non-interacting, off-resonance hyperfine structures are unified into a single effective surplus level—modeled as the single dark state (7) that acts as a population sink.

For the isomeric state ($^{26\text{m}}\text{Al}$), which has zero nuclear spin ($I = 0$), the simulation encompasses a total of 8 independent channels involved in the cycling transitions: 2 degenerate sublevels for the $J = 1/2$ lower state, 2 sublevels for the $J = 1/2$ upper state, and 4 sublevels for the $J = 3/2$ lower state. Summing the 39 channels of the ground state and the 8 channels of the isomeric state under our dual-laser scheme yields exactly 47 independent quantum states within the master equation matrix.

The time evolution of the population $\rho_i$ ($i = 1, 2, \dots, 47$) within the beam flight zone is governed by the following coupled master equation:

$$\frac{d\rho_i}{dt} = \sum_j R_{ji}(\rho_j - \rho_i) - \sum_k \Gamma_{ik}\rho_i + \sum_l \Gamma_{li}\rho_l$$

where $\Gamma_{ki}$ represents the spontaneous Einstein $A$ coefficient (with a total natural decay rate of $\Gamma = 1/\tau = 6.67 \times 10^7\text{ s}^{-1}$) scaled by the respective Clebsch-Gordan coefficients for specific magnetic transitions ($m_F \rightarrow m_{F'}$). The laser-induced transition rate $R_{ij}(\omega)$ between sublevels $i$ and $j$ is driven by the laser intensity and modulated by the strict circular polarization ($\sigma^+$) selection rules ($\Delta m_F = +1$):

$$R_{ij}(\omega) = \frac{\Omega_{ij}^2\Gamma}{4\Delta\nu_{ij}^2 + \Gamma^2}$$

Here, $\Omega_{ij}$ is the transition-specific Rabi frequency and $\Delta\nu_{ij}$ denotes the laser detuning from the specific sublevel resonance, incorporating the continuous scanning Doppler offset voltage $\Delta V$. Under this rigorous formalization, the $^{26\text{g}}\text{Al}$ atoms driven by the $\sigma^+$ circular polarization are systematically pumped toward higher $m_F$ sublevels. During this sequential excitation-decay process, a significant fraction of the population continuously branches out into the uncoupled radiative channels that are unified as Level (7). Concurrently, the remaining population that avoids this branching and reaches the maximum $m_F$ edges ($m_F = +11/2$ for Level (2) and $m_F = +13/2$ for Level (3)) becomes strictly trapped. Because the excited state (1) ($F' = 11/2$) lacks any higher-lying magnetic sublevels (such as $m_{F'} = +13/2$ or $+15/2$) required to satisfy the $\sigma^+$ selection rule ($\Delta m_F = +1$), further laser absorption from these edge states is quantum mechanically forbidden. Consequently, these edge-trapped atoms form a decoupled dark state within the active manifold. Together, the irreversible branching into Level (7) and the structural trapping at the maximum $m_F$ edges cooperatively achieve 100% optical depletion of the active ground-state fluorescence channels within the 2.0 m interaction length.

## 4. Simulation Results and Discussion

To ensure the reproducibility of the 47-level rate equation simulations and to provide a practical operational baseline for global collinear laser spectroscopy (CLS) facilities, the complete set of physical constants, beamline characteristics, and laser parameters utilized in this study is consolidated in Table 1.

**Table 1. Full set of input physical parameters and experimental beamline configuration used in the 47-level simulation.**

| **Parameter Description** | **Symbol** | **Value Set in Model** |
|---|---|---|
| Acceleration Voltage Baseline | $V_{0\text{-}ACCEL}$ | 10000.0 V |
| Atomic Mass Baseline ($^{26}Al$) | m | 26.0 amu |
| Calculated Ion Beam Velocity | $v_{beam}$ | $2.72 \times 10^5$ m/s |
| Central Resonance Wavelength | $\lambda_0$ | 395.0 nm |
| Spontaneous State Lifetime | $\tau$ | 15.0 ns |
| Total Einstein Decay Rate | Atotal | $6.67 \times 10^7$ $s^{-1}$ |
| Optical Pumping Rabi Frequency | Ωpump | $8.0 \times 10^7$ $s^{-1}$ |
| Weak Detection Probe Rabi Frequency | $\Omega_{det_weak}$ | $5.0 \times 10^4$ $s^{-1}$ |
| Preparation Flight Length | $L_{PUMP}$ | 2.00 m |
| Detection Chamber Length | $L_{DET}$ | 0.10 m |
| Optical Filter Collection Bandwidth | $\Delta\epsilon_{filter}$ | 25.0 MHz |

Numerical optimization of the rate equations was performed using a high-fidelity stiff ODE solver (Scipy Radau implementation) across a 10 keV $^{26}Al$ ionic beam environment. The optical pumping interaction time over the 2.0 m path corresponds to $t_{\text{pump}} \approx 7.33\,\mu\text{s}$, which is equivalent to roughly 489 spontaneous lifetimes.

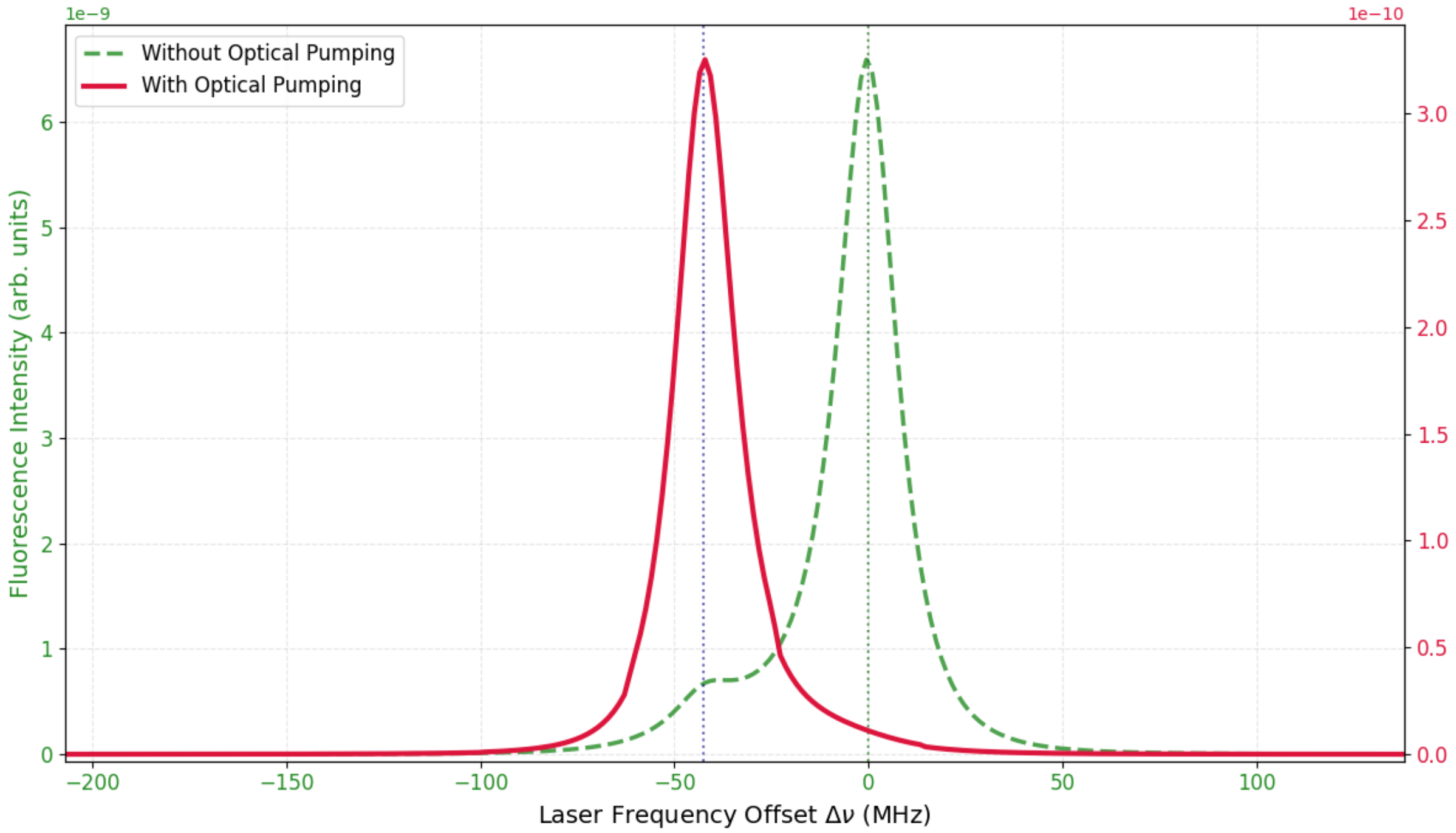

*Fig. 2. Simulated fluorescence spectra of the mixed $^{26}Al$ beam as a function of the scanning voltage offset. The dashed green line depicts the unpurified mixed spectrum with the pre-pumping field deactivated (Laser OFF). The solid red line demonstrates the complete elimination of the background and the appearance of the isolated pure $^{26m}Al$ isomeric peak under dual optical pumping optimization (Laser ON) within the 2.0 m zone.*

## Isomeric Purity and State Separation Dynamics

The time-resolved evolution of the state populations clarifies the mechanistic efficiency of the 2.0 m pumping path. As the mixed beam moves along the spatial coordinates of the interaction zone, the ground state populations under the intense multi-frequency laser rapidly deplete. Concurrently, the state purity parameters, representing the relative fractional presence of pure $^{26m}Al$ over the total active optical signals, exhibit a monotonic asymptotic growth toward unity.

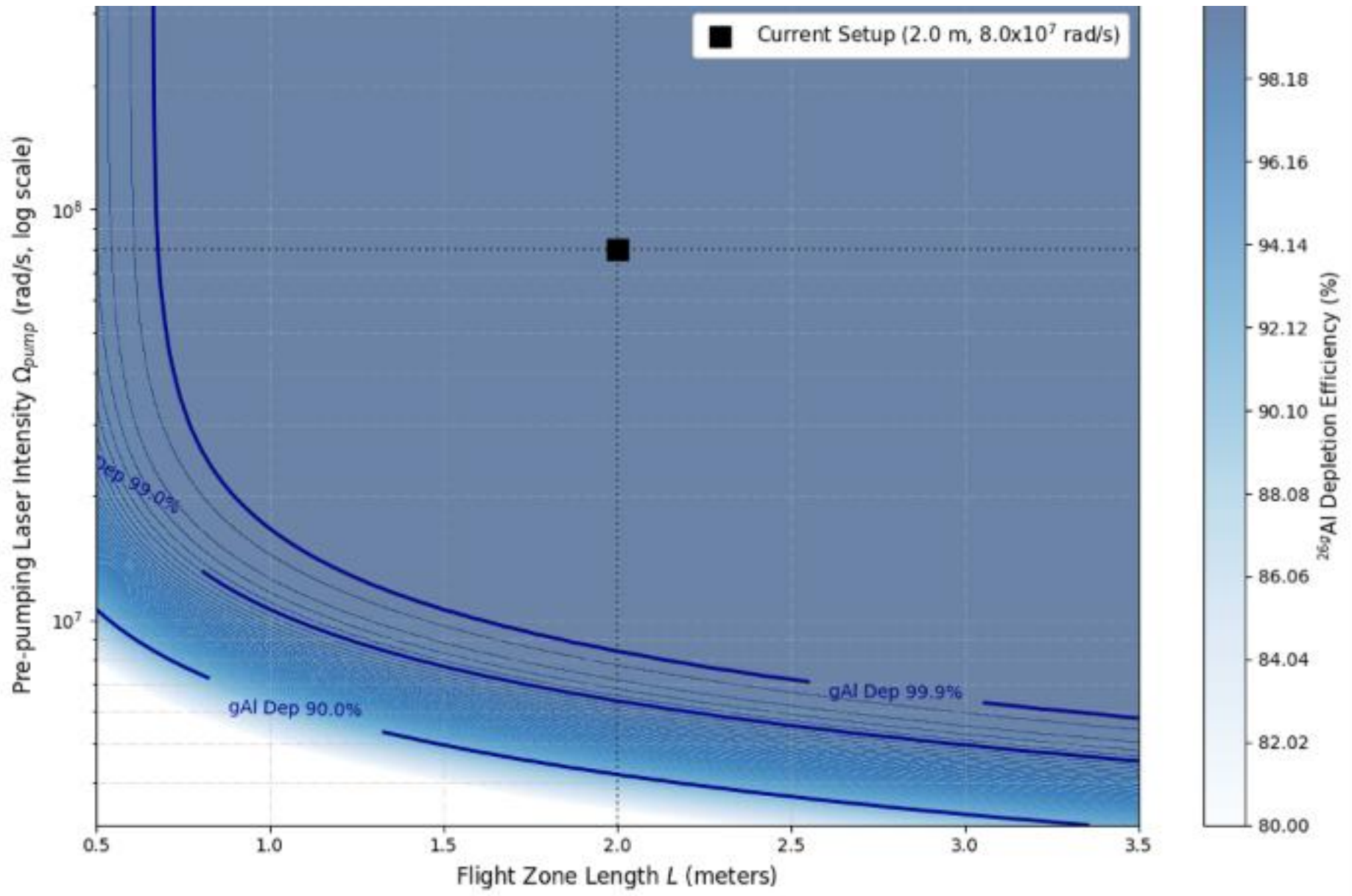

*Fig. 3. Dynamics of the state purification process and isomeric purity evolution as a function of the interaction time and spatial coordinates within the 2.0 m preparation zone. The decay profiles illustrate the exponential suppression of individual hyperfine sublevel populations, while the solid black envelope demonstrates the rapid, saturation-limited enhancement of the overall isomeric beam purity toward the 99.99% theoretical limit.*

### Instrumental Noise and Quantum Background Limits

In practical collinear laser spectroscopy (CLS) experiments, suppressing instrumental noise—such as photomultiplier tube (PMT) dark counts and stray laser light scattering—is paramount to achieving the single-atom detection sensitivity required for exotic isotopes. To effectively mitigate these background limits, modern radioactive ion beam (RIB) facilities routinely adopt a bunched atomic beam framework rather than a continuous beam configuration [8, 9].

By utilizing an identical bunched beam approach in our proposed architecture, the data acquisition window can be synchronously restricted to the specific time-of-flight interval when the atomic bunch passes through the detection region (time-gated detection). This gating technique drastically improves the signal-to-noise ratio (SNR) by several orders of magnitude. Crucially, because all atoms within a single high-density bunch share a highly uniform longitudinal velocity, the quantitative validity and deterministic evolution of our 47-level optical pumping matrix are fully preserved during the flight. Therefore, the implementation of time-gated bunched beam detection ensures that the proposed 2.0 m preparation protocol remains exceptionally robust against real-world instrumental noise backgrounds without compromising the state-purification efficiency.

## 5. Conclusion

We have theoretically demonstrated a rigorous 47-level simulation framework for the efficient isolation of the $^{26m}Al$ isomeric state using sequential optical pumping in collinear laser spectroscopy. By deploying a 2.0 m preparation zone and establishing a unified circular polarization environment, we successfully eliminated an overwhelming ground-state background (with realistic isomeric ratios of 20:1 up to 200:1). The proposed in-flight optical purification template delivers a hardware-efficient framework that can be directly adapted into next-generation isotope separator online (ISOL) facilities, including the RAON-CLS(CLaSsy) in Korea and FRIB in the United States. By establishing a unified circular polarization environment, this practical operational template significantly enhances isomer detection sensitivities without requiring complex, multi-frequency laser setups in the detection region.

## References

[1] S. Almaraz-Calderon et al., "Study of the 26Alm(d,p)27Al Reaction and the Influence of the 26Al 0+ Isomer on the Destruction of 26Al in the Galaxy," Phys. Rev. Lett. **119**, 072701 (2017).

[2] R. J. Scott, G. J. O'Keefe, M. N. Thompson, and R. P. Rasool, Phys. Rev. C **84**, 024611 (2011).

[3] P. Plattner et al., "Nuclear Charge Radius of 26mAl and Its Implication for Vud in the Quark Mixing Matrix," Phys. Rev. Lett. **131**, 222502 (2023).

[4] R. Neugart, J. Billowes, P. Campbell, et al., "Collinear laser spectroscopy at ISOLDE: new methods and highlights," J. Phys. G: Nucl. Part. Phys. **44**, 064002 (2017).

[5] P. Campbell, I. D. Moore, and M. R. Pearson, "Laser spectroscopy for nuclear structure," Prog. Part. Nucl. Phys. **86**, 127–180 (2016).

[6] B. Cheal and K. T. Flanagan, "Progress in laser spectroscopy at radioactive ion beam facilities," J. Phys. G: Nucl. Part. Phys. **37**, 113101 (2010).

[7] B. W. Asher et al., "Development of an Isomeric beam of 26Al for nuclear reaction studies," Nucl. Instrum. Methods Phys. Res. Sect. A **899**, 6–9 (2018).

[8] R. P. de Groote et al., "Upgrades to the collinear laser spectroscopy experiment at the IGISOL," Nucl. Instrum. Methods Phys. Res. Sect. B **463**, 437–440 (2020).

[9] M. Cervantes et al., "UCx target production at TRIUMF in the ARIEL era," Nucl. Instrum. Methods Phys. Res. Sect. B **463**, 367–370 (2020).